\documentclass[12pt]{iopart}

\usepackage{graphicx}
\usepackage{graphics}
\usepackage{epsfig}
\begin{document}

\title[Prospects for $\phi$ meson production in pp collisions at the ALICE experiment]{Prospects for $\phi$ meson production in pp collisions at the ALICE experiment}

\author{J.D. Tapia Takaki (for the ALICE Collaboration)}

\address{Particle Physics Group, School of Physics and Astronomy, \\
The University of Birmingham, Edgbaston, Birmingham B15 2TT, UK}
\ead{jdtt@hep.ph.bham.ac.uk}
\begin{abstract}
The ALICE experiment at the CERN Large Hadron Collider (LHC) will allow the
study of resonance production in nucleus-nucleus and proton-proton collisions.
This paper presents results based on physics performance studies to discuss prospects 
in ALICE for $\phi$(1020) meson production in pp interactions during the LHC startup.
\end{abstract}
\maketitle

%
\section{Introduction to resonance production}
%

ALICE is a general-purpose heavy ion experiment designed to
study the physics of strongly interacting matter and the 
Quark Gluon Plasma (QGP) in nucleus-nucleus collisions 
at the LHC~\cite{Alessandro:2006yt}. Resonances, in particular the $\phi$ meson, 
are useful probes to study the high density medium created in ultrarelativistic 
heavy ion collisions. The hadronic cross section associated with the $\phi$ meson 
is small, which makes this particle rather insensitive to the presence of other hadrons 
in the late stage of the collision. Therefore, the production of $\phi$ mesons has been 
suggested as a signature for strangeness production mechanisms owing to an early partonic 
phase~\cite{Rafelski:1982pu,Koch:1986ud,Bass:1999zq}. Preliminary results from the STAR
collaboration suggest that such mechanisms might have been observed at RHIC energies when 
comparing the $\phi$ yields obtained in Au-Au collisions, and more recently in Cu-Cu collisions,
with those obtained in pp collisions~\cite{nxusqm07}. Although the lifetime of the $\phi$ meson
in vacuum is larger than that expected for the QGP state, significant medium modifications of its
spectral properties have been predicted~\cite{Klingl:1997tm, Rapp:1999ej}. As a consequence, the 
branching ratio for its decay into kaon and lepton pairs may change. The observation of such 
modifications might also provide information on the mechanism relevant for $\phi$ production in 
high energy collisions, which at present remains an open question.

In addition, the $\phi$ meson is an interesting particle in itself as a hadronic measurement
that could be done at several energies and for various collision systems at the LHC.
Analysing resonances in pp collisions is important as a benchmark for the heavy ion run. 
Moreover, it is also a significant analysis as $\phi$ production has not been measured to 
very high precision at the Tevatron~\cite{Alexopoulos:1995ru}, so even ``low energy" points 
from the LHC startup would become the best data at that energy.

Furthermore, $\phi$ mesons could be used as indicators of strangeness production along with the particles with
open strangeness ($K^{\pm},K^{0},\Lambda,\Xi, \Omega$)~\cite{VillalobosBaillie:2005nk}. It indicates the level of strangeness production as there
are predictions that the strange sea could be large~\cite{Ellis:2000kh}. At the LHC it would be possible to access the very
low-x region, about which not much is known. Additionally, HERA measurements will need verification
and improvements~\cite{Chekanov:2005cqa}.

%
\section{Resonance detection in ALICE}
\begin{figure}
\begin{center}
\includegraphics*[scale=0.6]{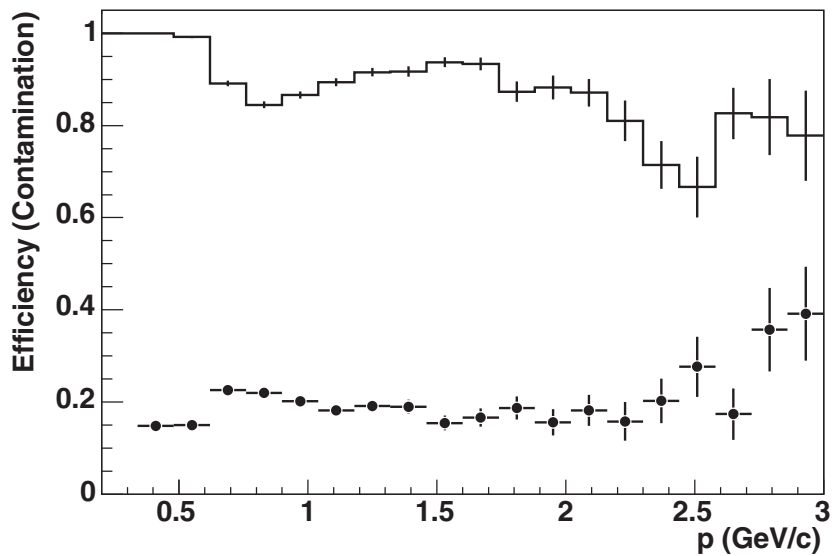}
\caption{Combined efficiency (solid line) and contamination (points with error bars) for 
charged kaon identification using the ITS, TPC and TOF detectors for PbPb collisions~\cite{Alessandro:2006yt}.}
\label{fig:pidkaons}
\end{center}
\end{figure}

%
A key feature of the ALICE experiment is its very good capability to identify charged and some neutral 
particles using a variety of detector techniques, which can resolve the different particle ambiguities 
at different momentum ranges. ALICE can combine the PID information from single detectors to cover the 
different kinematic limits. Table~\ref{pidkaonstable} shows the momentum range over which 
kaons can be identified using the specified detector and technique. In particular, the TPC and ITS, 
which give dE/dx measurements, covering the full central region and can be used to identify charged 
particles from below 1 GeV/{\it c}. The TOF is used for hadron identification by Time of Flight. The efficiency 
and contamination for kaons is shown in Figure~\ref{fig:pidkaons}. The high momentum particle identification
(HMPID) system was not included in this Figure but can be used for the identifications of kaons in the medium 
range of momentum at a limited solid angle coverage. In addition, dE/dx measurements can be used again to identify 
kaons at high momenta using the relativistic rise phenomenon~\cite{Alessandro:2006yt}.

\begin{table}
\caption{\label{math-tab2}Momentum range over which kaons can be identified using the
specified detector and technique. The mid-rapidity range (-0.9$<\eta<$0.9) is only considered~\cite{Alessandro:2006yt}.}
\begin{tabular*}{\textwidth}{@{}l*{15}{@{\extracolsep{0pt plus
12pt}}l}}
\br
Range (GeV/{\it c}) & PID technique/sub-detector\\
\mr
0.1-0.5 & dE/dx (ITS+TPC)\\
0.35-2.5 & Time of Flight \\
1-3 & HMPID \\
5-50 & Relativistic rise (ITS+TPC) \\
0.3-13 & Secondary vertex reconstruction\\
\br
\label{pidkaonstable}
\end{tabular*}
\end{table}

%
\section{Expectations for $\phi$ meson production and backgrounds}
\begin{figure}
\begin{center}
\includegraphics[width=0.45\textwidth]{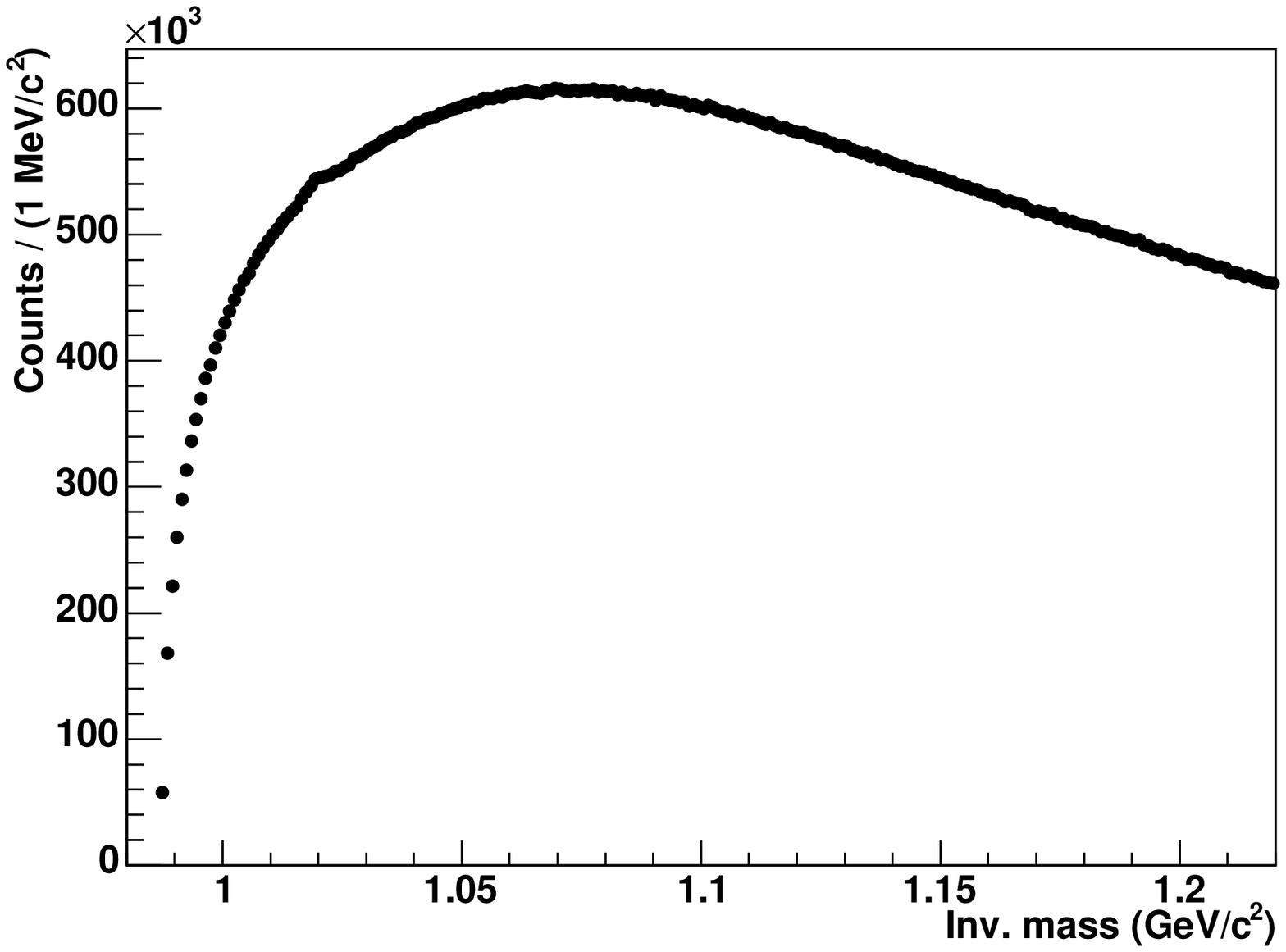}
\includegraphics[width=0.45\textwidth]{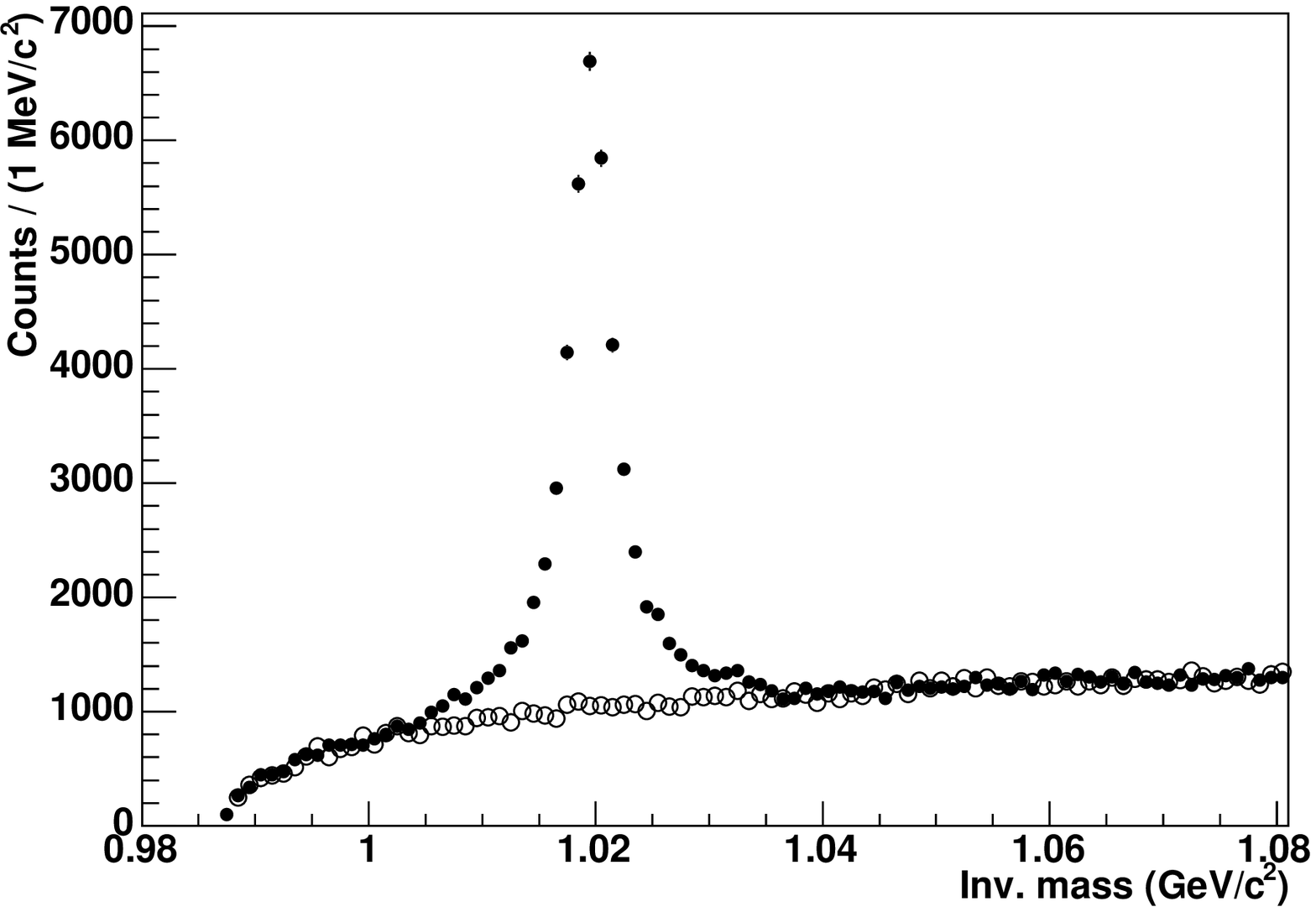}
\caption{Predictions for the $K^{+}K^{-}$ invariant mass spectrum for pp collisions: (Left) without
making use of the ALICE PID capabilities; (Right) assuming perfect particle identification efficiency 
for charged kaons. The background was estimated by like-sign method (open points).}
\label{fig:phiInvMass}
\end{center}
\end{figure}

%
Apart from the identification of charged particles, resonance studies require a good understanding
of the combinatorial background. Figure~\ref{fig:phiInvMass} (left) shows the invariant mass distribution
of kaon pairs without using the PID system. A sample of 7x$10^{6}$ minimum bias PYTHIA pp events at $\sqrt{s}$= 14 TeV
was analysed (same statistics for the rest of plots). The sample was divided into three multiplicity ranges.
The figures shown in this paper correspond to the (intermediate) range between 5 and 25 in the negative charged track 
multiplicity at mid-rapidity. Mixing particles from different events has been a technique used in the past to estimate 
the combinatorial background. The limitation of this method is that some ``event similarity" condition needs 
to be imposed, which strongly depends on the multiplicities presented in the event. Combining particles 
from the same event is also another option, using like-sign pairs. Both methods were implemented in 
this analysis, providing similar results as was previously reported in~\cite{Adler:2004hv}. These techniques are 
currently being optimised to make use of distributed events by GRID services where the final analysis is foreseen. 

The selections for kaon tracks have yet to be finalised for the first analysis with real data, but some preliminary 
results will be presented~\cite{takaki:2007jk}. The simulation of pp interactions is generated by PYTHIA 6.214 Monte Carlo
event generator. The output is then passed to the ALICE simulation and reconstruction software using the standard computing framework. 
Results presented here are based on a fast-simulation method described in detail in~\cite{takaki:2007jk}. Because
the short lifetime of resonances, they all decay at the primary vertex, so the track selection exclude particles from secondary 
interactions. All tracks are required to come from the primary vertex by using impact parameter cuts. Figure~\ref{fig:phiInvMass} (right) shows the invariant mass distribution and the estimation of the background using the like-sign method and assuming perfect PID efficiency for kaons. The reconstructed mass and width, as well as the number of $\phi$ mesons after the background subtraction, were found to be 
consistent with the generated values. A similar study, using the full set of ALICE detectors to simulate realistic conditions for PID, 
leads to mass spectra (not shown here) very similar to those in Figure~\ref{fig:phiInvMass} (right). More details are given in Reference~\cite{takaki:2007jk}.

%
\section{$\phi$ meson detection during the LHC startup}
\begin{figure}
\begin{center}
\includegraphics[width=0.45\textwidth]{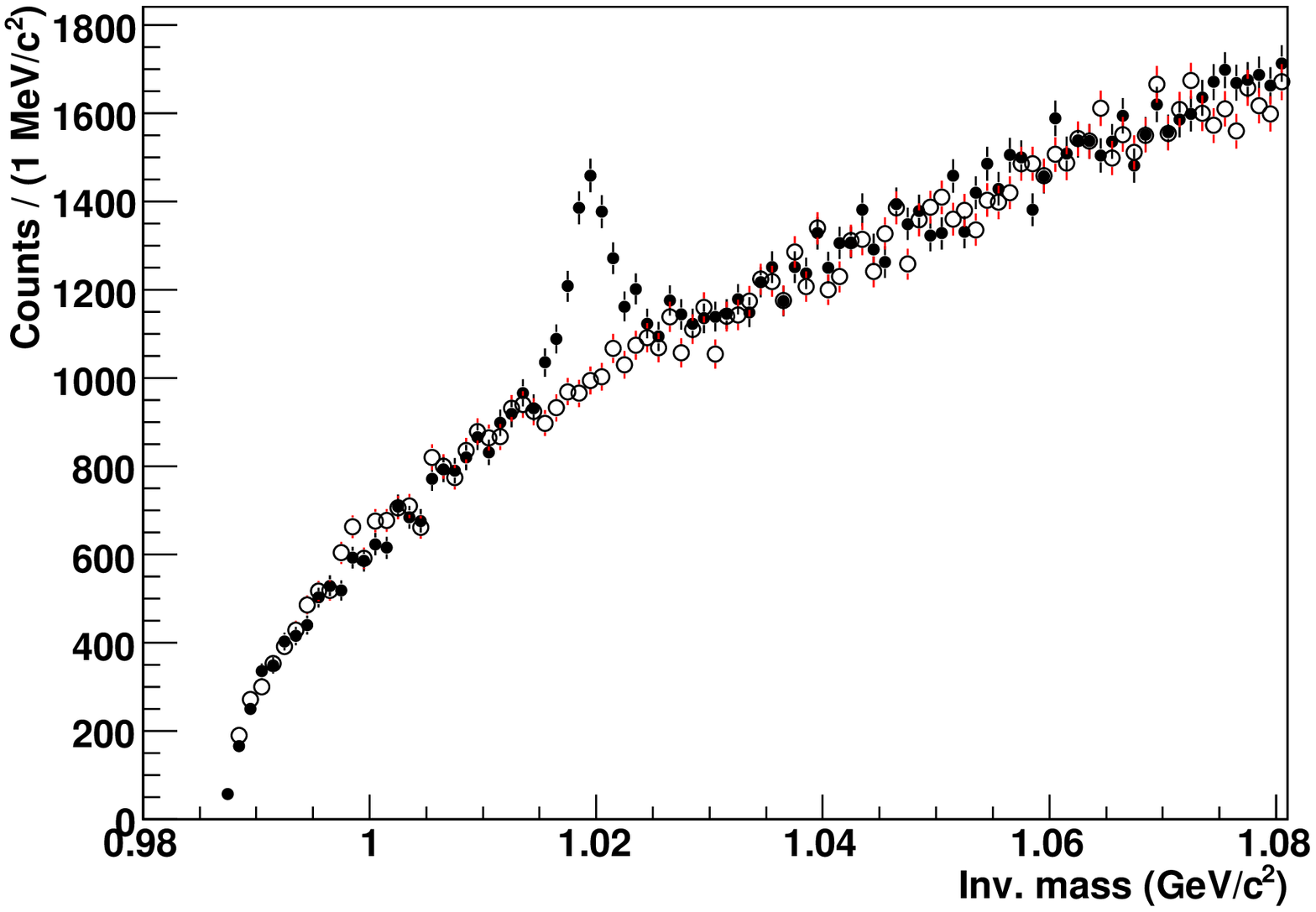}
\includegraphics[width=0.45\textwidth]{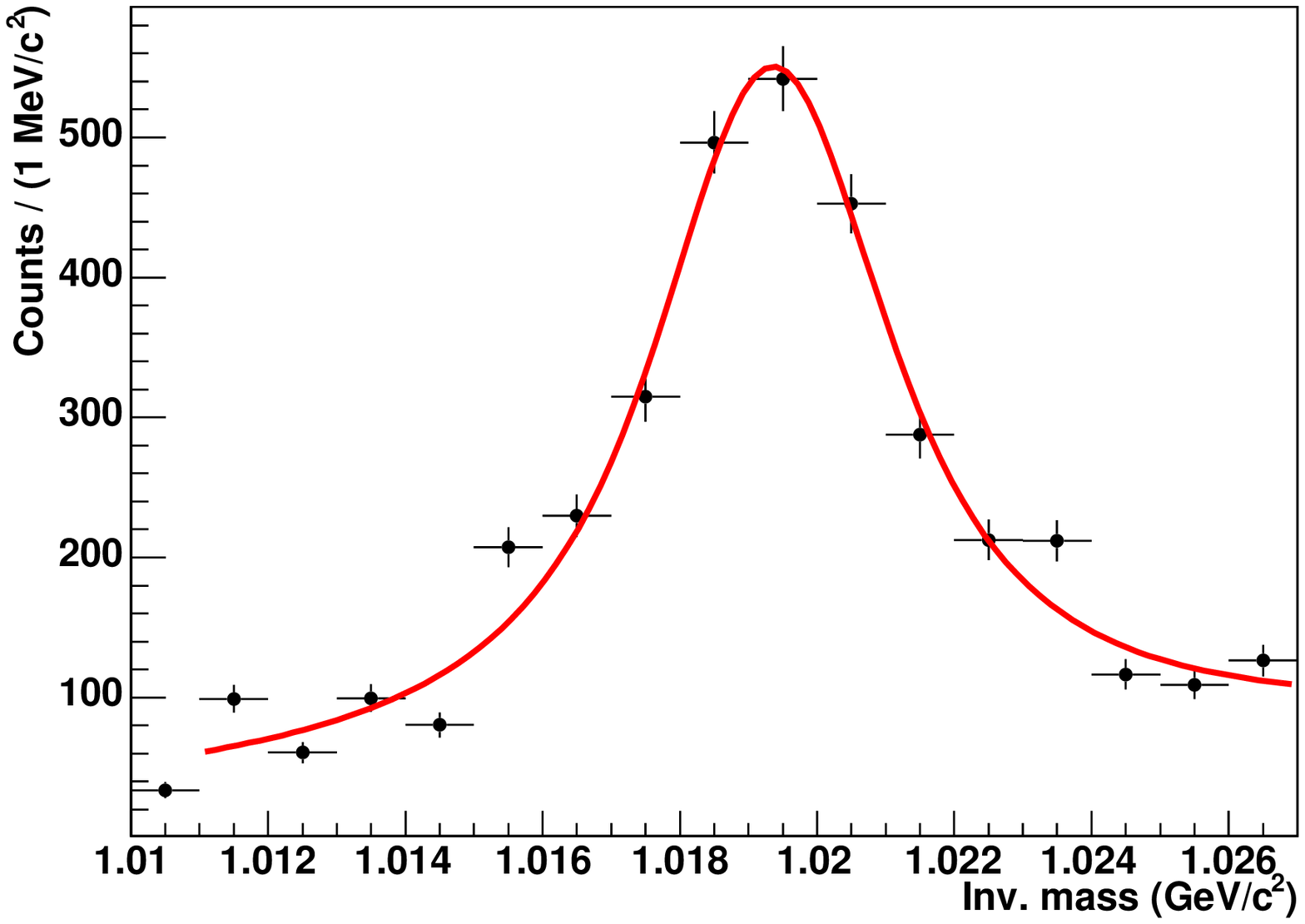}
\caption{Predictions for the $K^{+}K^{-}$ invariant mass spectrum for pp collisions without accessing the PID
information from the detectors (left) and after the subtraction of the background (right). The background was estimated
by like-sign method (open points). A cut on the transverse momentum of the pair of 2.4$<p_{t}<$2.8 GeV/{\it c}  was used.}
\label{fig:phinopid}
\end{center}
\end{figure}

%
Because of the narrow width and good signal-to-background ratio for the $\phi$, ALICE should be able to detect this resonance
early in the running of LHC. However, during that period the particle identification system might not be fully 
available. A study was carried out to investigate the prospects for measuring $\phi$ mesons without accessing any 
of the PID information provided by the detectors. It was found that by applying a relatively high momentum cut 
on the candidate pair, the $\phi$ meson peak can only be resolved to an acceptable level. Figure~\ref{fig:phinopid} 
shows the predictions for the $K^{+}K^{-}$ invariant mass spectrum for pp collisions for pairs with $2.4<p_{t}<2.8$ GeV/{\it c} 
(left hand side) and after the subtraction of the background (right), showing that a resonance signal above the combinatorial 
background can be obtained from 7x$10^{6}$ events.

%
\section{Summary}
%
The ALICE experiment will allow the study of resonance production in great detail due
to its very good particle identification and tracking capabilities. A study of the $\phi$ meson in hadronic (and leptonic)
decay channels will be possible. Like-sign and mixing-event methods were tested
for background subtraction. They give consistent results. The $\phi$ meson analysis 
might be one of the first resonance measurements in ALICE because of the good signal-to-background 
ratio. ALICE should be able to study the $\phi$ meson in pp collisions before the PID system becomes 
available. Results presented here indicate that it will be possible to measure the mass distribution
and momentum spectra for the $\phi$ meson through its hadronic decay $\phi\rightarrow K^{+}K^{-}$ 
during the ``first physics" days at the LHC.

\ack
This work was supported by the UK Overseas Research Students Awards Scheme (ORSAS)
and the University of Birmingham, and by the EU Al$\beta$an programme.

\end{document}